\documentclass[a4paper]{article}
\usepackage[utf8]{inputenc}
\usepackage{verbatim}
\usepackage{float}
\usepackage{mathtools}
\usepackage{multirow}
\usepackage{amsmath}
\usepackage{amssymb}
\usepackage{graphicx}

\makeatletter

\pdfpageheight\paperheight
\pdfpagewidth\paperwidth

\providecommand{\tabularnewline}{\\}

\pdfoutput=1
\usepackage{jheppub}
\usepackage{amsfonts}
\usepackage{relsize}
\usepackage{pdfpages}
\usepackage{multicol}
\usepackage{pdfpages}
\usepackage{epstopdf}
\usepackage{caption}

	\title{Power corrections to event shapes using Eikonal
Dressed Gluon Exponentiation
} 
	
	\author[a]{Neelima Agarwal,}
	\author[b]{Ayan Mukhopadhyay,}
	\author[b]{Sourav Pal,}
	\author[b]{Anurag Tripathi,}
	\affiliation[a]{Department of Physics, Chaitanya Bharathi Institute of Technology, \\ Gandipet, Hyderabad, Telangana State 500075, India}
	\affiliation[b]{Department of Physics, Indian Institute of Technology Hyderabad, \\ Kandi, Sangareddy, Telangana State 502285, India}
	
	\emailAdd{neelimaagarwal$\_$physics@cbit.ac.in}
	\emailAdd{ayanmukhopadhyay5@gmail.com}
	\emailAdd{spalexam@gmail.com}
	\emailAdd{tripathi@phy.iith.ac.in}

\abstract{Event shapes are classical tools for the determination of the strong coupling and for
the study of hadronization effects in electron-positron annihilation. In the context of analytical
studies, hadronization corrections take the form of power-suppressed contributions to the cross
section, which can be extracted from the perturbative ambiguity of Borel-resummed distributions.
We propose a simplified version of the well-established method of Dressed Gluon Exponentiation
(DGE), which we call Eikonal DGE (EDGE), which determines all dominant power corrections
to event shapes by means of strikingly elementary calculations. We believe our method can be
generalized to hadronic event shapes and jet shapes of relevance for LHC physics.}

\makeatother

\begin{document}
\maketitle

\section{Introduction}

\label{intro} ~ Infrared-safe event shape variables, which we will
generically denote by $e$, play a central role in perturbative QCD:
they are essential tools for the precise determination of the strong
coupling constant, and they are classic testing grounds for both analytical
and numerical models of hadronization. Owing to their infrared and
collinear safety, they can be computed in perturbation theory, and
furthermore large logarithmic corrections to the distributions can
be resummed to all orders by a variety of methods. At fixed orders,
the state of the art is next-to-next-to-leading order (NNLO) accuracy
\cite{GehrmannDeRidder:2007hr,Weinzierl:2009ms,Gehrmann:2019hwf,Kardos:2020igb,GehrmannDeRidder:2009dp},
while the next to leading log (NLL) resummation has been known for
a while~\cite{Catani:1992ua,Banfi:2010xy,Banfi:2004yd,Banfi:2001bz}.
In recent years, the NNLL resummation framework has also been developed
\cite{Banfi:2016zlc,Abbate:2010xh,Becher:2012qc,Hoang:2014wka,Budhraja:2019mcz,Bell:2018gce,Hoang:2015gta,Kolodrubetz:2016reu,Lepenik:2019jjk,Lee:2009cw,Hornig:2009vb,Banfi:2014sua}).
Here we will be concerned with analytic estimates of non-perturbative
corrections, which are suppressed by powers of $\Lambda/Q$ (where
$\Lambda$ is the QCD scale and $Q$ is the center-of-mass energy)
with respect to the perturbative result. The basic idea of such analytic
estimates goes back to the Operator Product Expansion (OPE), and was
first applied to observable that do not admit an OPE in the early
papers~\cite{Bigi:1994em,Beneke:1994sw,Webber:1994cp}. Very roughly
speaking, one notes that a generic (dimensionless) observable in perturbative
QCD is a sum of a `leading power' perturbative series plus power corrections,
of the general form 
\begin{equation}
\sigma\left(\frac{Q}{\mu},\alpha_{s}\right)\,=\,\sigma_{{\rm pert}}\left(\frac{Q}{\mu_{f}},\frac{\mu_{f}}{\mu},\alpha_{s}\right)+\sum_{n}\sigma_{n}\left(\frac{\mu_{f}}{\mu},\alpha_{s}\right)\left(\frac{\mu_{f}}{Q}\right)^{n}\,,\label{gensig}
\end{equation}
where $\mu_{f}$ is a perturbative factorization scale, ultimately
to be traded for the strong interaction scale $\Lambda$. Generically,
with a dimensional regulator, different terms in the sum in Eq.~(\ref{gensig})
mix with each other under renormalization. In dimensional regularization,
the same effect arises in a subtler fashion: each term in Eq.~(\ref{gensig})
is ambiguous due to the divergence of the corresponding perturbative
expansion, which manifests itself via singularities in the Borel plane.
These ambiguities are power-suppressed and are compensated by corresponding
ambiguities in subsequent terms in the sum in Eq.~(\ref{gensig}).
This opens the way for a perturbative estimate of hadronization corrections
based on the study of singularities in the Borel plane. Phenomenological
studies of event shapes and other basic QCD observables with these
tools were first systematically pursued in~\cite{Dokshitzer:1995qm},
and subsequently developed in a vast literature, reviewed in~\cite{Dasgupta:2003iq}.
The phenomenological importance of these power-suppressed corrections
cannot be understated: for example, they are crucial for a precise
determination of the strong coupling~\cite{Abbate:2010xh,Hoang:2015hka,Wang:2019isi,Marzani:2019evv,Gehrmann:2012sc}.

In the case of event shape distributions, denoted by $d\sigma/de$
below, the situation is more subtle. Such distributions peak in the
two-jet region, which can be taken to correspond to $e\to0$, and
which is dominated by soft and collinear emissions; in this region,
the distributions are typically affected by enhanced power corrections
of the form $\left(\Lambda/(eQ)\right)^{n}$, associated with the
emission of soft gluons, as well as corrections scaling as $\left(\Lambda^{2}/(eQ^{2})\right)^{n}$,
associated with hard collinear gluon emission. We will refer to the
first of these as `soft' power corrections, and to the second ones
as `collinear' power corrections for the sake of brevity. When $e\sim\Lambda/Q$,
which is typically close to the peak of the distribution at least
at LEP energies, all soft power corrections become equally important
and need to be resummed in order to get a stable prediction. At even
smaller values of $e$, $e\sim\Lambda^{2}/Q^{2}$, collinear power
corrections become relevant as well.

An elegant and efficient method to handle simultaneously large perturbative
logarithms (up to NLL accuracy) and power corrections in the two-jet
region is Dressed Gluon Exponentiation (DGE)~\cite{Gardi:2001di},
which has already been applied to a variety of event shapes~\cite{Gardi:2002bg,Gardi:2003iv,Berger:2004xf},
as well as to other important QCD observables~\cite{Cacciari:2002xb,Gardi:2004ia,Andersen:2005bj}.
DGE, aside from consistently including the NLL resummation of Sudakov
logarithms, provides a renormalon-based estimate of both soft and
collinear power corrections. Collinear power corrections have been
shown to enjoy a degree of universality~\cite{Gardi:2002bg,Gardi:2003iv,Gardi:2004ia}
across several inclusive observables. When, however, this universality
breaks down, as for example in~\cite{Berger:2004xf,Lee:2007jr},
collinear power corrections can be very cumbersome to compute; furthermore,
they only become relevant at extremely small values of the event shape,
usually out of experimental range, or in a region where very few data
points are available.

These facts suggest that it would be useful to construct a systematic
approximation to DGE which would suffice to capture all soft power
corrections, while remaining simple to implement in practice. In this
paper, we will introduce such an approximation, which essentially
consists in combining DGE with the eikonal approximation for the relevant
matrix elements. We call the resulting construction Eikonal Dressed
gluon exponentiation or EDGE. The universality and simplicity of soft
emission can then be used to express soft power corrections to a large
class of event shapes in terms of a very simple integral, which reproduces
known results for all event shapes for which soft power corrections
are known. The paper is structured as follows: section 2 briefly summarizes
the essential aspects of DGE, section 3 shows how to implement EDGE
using energy fractions by taking examples from three very well known
event shapes: thrust, $C$-parameter and angularities, section 4 describes
the implementation of EDGE using the transverse momentum and the rapidity,
in section 5 we present Sudakov exponent and discuss power corrections.

\section{Dressed Gluon Exponentiation}
\label{DGE}

The starting point for DGE is the event shape distribution in the
single dressed gluon approximation, which is constructed from the
one-loop real emission contribution to the event shape for a gluon
with virtuality $k^{2}\neq0$. From this, one easily obtains~\cite{Beneke:1994sw}
the (renormalon) resummation of quark vacuum polarization corrections
which dominates in the large $N_{f}$ limit. One can write the result
as 
\begin{equation}
\frac{1}{\sigma}\frac{d\sigma}{de}(e,Q^{2}) \Big|_{SDG} \,=\,-\,\frac{C_{F}}{2\beta_{0}}\int_{0}^{1}d\xi\,\frac{d\ensuremath{\mathcal{F}}(e,\xi)}{d\xi}\,A(\xi Q^{2})\,,\label{SDG}
\end{equation}
where $\beta_{0}=\frac{11}{12}C_{A}-\frac{1}{6}N_{f}$, $\xi=k^{2}/Q^{2}$,
and $A(\xi Q^{2})$ is the large-$\beta_{0}$ running coupling $(A=\beta_{0}\alpha_{s}/\pi)$
on the time-like axis. In the $\overline{{\rm MS}}$ scheme, it admits
the Borel representation 
\begin{equation}
A(\xi Q^{2})\,=\,\int_{0}^{\infty}du(Q^{2}/\Lambda^{2})^{-u}\frac{\sin\pi u}{\pi u}{\rm e}^{\frac{5}{3}u}\xi^{-u}\,.\label{largeb0coup}
\end{equation}
The cornerstone of Eq.~(\ref{SDG}) is the characteristic function
$\ensuremath{\mathcal{F}}(e,\xi)$, which is the one-loop event shape
distribution with a non-vanishing gluon virtuality $k^{2}$~\cite{Gardi:2000yh,Dokshitzer:1995qm},
\begin{equation}
\ensuremath{\mathcal{F}}(e,\xi)\,=\,\int dx_{1}dx_{2}\,{\cal M}(x_{1},x_{2},\xi)\,\delta\left(e-\bar{e}(x_{1},x_{2},\xi)\right)\,,\label{charf}
\end{equation}
where $x_{i}$ are the customary energy fraction variables, ${\cal M}$
is the matrix element for the emission of a gluon with $k^{2}\neq0$,
and $\bar{e}$ is the explicit expression of the event shape in terms
of the kinematic variables. Interchanging the order of integrations
in Eq.~(\ref{SDG}) we can construct a Borel representation as 
\begin{equation}
\frac{1}{\sigma}\frac{d\sigma}{de}(e,Q^{2}) \Big|_{SDG}\,=\,\frac{C_{F}}{2\beta_{0}}\int_{0}^{\infty}du(Q^{2}/\Lambda^{2})^{-u}B(e,u)\,,\label{borrep}
\end{equation}
where the Borel function $B(e,u)$ is defined by 
\begin{equation}
B(e,u)\,=\,-\frac{\sin{\pi u}}{\pi u}e^{\frac{5}{3}u}\int_{0}^{\infty}d\xi~\xi^{-u}\frac{d\ensuremath{\mathcal{F}}(e,\xi)}{d\xi}\,.\label{borfu}
\end{equation}
The Borel function $B(e,u)$ has a simple structure in the $u$ plane,
without renormalon singularities. Renormalon poles are however generated
when the single dressed gluon distribution is exponentiated via a
Laplace transform~\cite{Gardi:2001di}.

The additive property of the event shapes with respect to the multiple
gluon emissions together with the factorization of soft and collinear emissions
from the {\it hard} part of the matrix element leads to the
exponentiation of the logarithmically 
enhanced terms in the Laplace space and the resummed cross section
is given by \cite{Gardi:2002bg,Gardi:2003iv}, 
\begin{align}
\frac{1}{\sigma}\frac{d\sigma(e,Q^{2})}{de}=\int_{C-i\infty}^{C+ i\infty}\frac{d\nu}{2\pi i}\,e^{\nu e}\,\text{exp}[S(\nu,Q^{2})] & ,\label{eq:resum}
\end{align}
where $C$ lies to the right of the singularities of the integrand.
The Sudakov exponent has the form \cite{Gardi:2001ny}, 
\begin{align}
S(\nu,Q^{2}) & =\int_{0}^{1}de\:\frac{1}{\sigma}\frac{d\sigma(e,Q^{2})}{de} \Big|_{SDG} \:(e^{-\nu e}-1).\label{eq: sudakov-gen}
\end{align}
The Sudakov region $e\rightarrow0$ corresponds to $\nu\rightarrow\infty$. Using Eq. (\ref{borrep}), the
Sudakov exponent takes the form
\begin{align}
S(\nu,Q^{2})=\frac{C_{F}}{2\beta_{0}}\int_{0}^{\infty}du\,\bigg(\frac{Q^{2}}{\Lambda^{2}}\bigg)^{-u}\,B_{\nu}^{e}(u),\label{eq:sudakov-gen-1}
\end{align}
where the Borel function in the  Laplace space, $B_{\nu}^{e}(u)$, is defined as 
\begin{align}
B_{\nu}^{e}(u)=\int_{0}^{1}de\,B(e,u)\,(e^{-\nu e}-1).\label{eq: borel-int}
\end{align}
This exponentiation effectively resums both large Sudakov logarithms
and power corrections in the two-jet region.
\section{Borel function using Eikonal Dressed Gluon Exponentiation}

\label{EDGEenergy}

In this article we undertake the calculation of the Borel function that was defined in Eq.~(\ref{borfu})  for
three very well known event shape variables: (a) the thrust~\cite{Brandt:1964sa,Kane:1978nd,Farhi:1977sg,Altarelli:1981ax},
(b) the $C$-parameter~\cite{Donoghue:1979vi,Fox:1978vu,Bjorken:1969wi,Parisi:1978eg}
and, (c) the angularities~\cite{Berger:2003pk,Berger:2003iw,Berger:2002ig},
and we propose a simplified version of the well-established method
of Dressed Gluon Exponentiation (DGE), which we call Eikonal DGE (EDGE),
which determines all dominant power corrections to event shapes by
means of strikingly elementary calculations. We believe our method
can be generalized to hadronic event shapes and jet shapes of relevance
for LHC physics. There are two aspects to this simplification. First,
as we will see in the later parts of this article, we only need to
work with the squared matrix element in the eikonal limit. Second,
and more importantly, the event shape definitions can be simplified
(\textit{eikonalized}) to significantly simplify the computations, however,
still capturing the dominant power corrections. The definition of
thrust is simple enough and does not require any \textit{eikonalization}, however
we will introduce the  \textit{eikonalized} versions of $C$-parameter and
angularities for the computation of their respective Borel functions.

\begin{figure}[b]
\centering \includegraphics[scale=0.6]{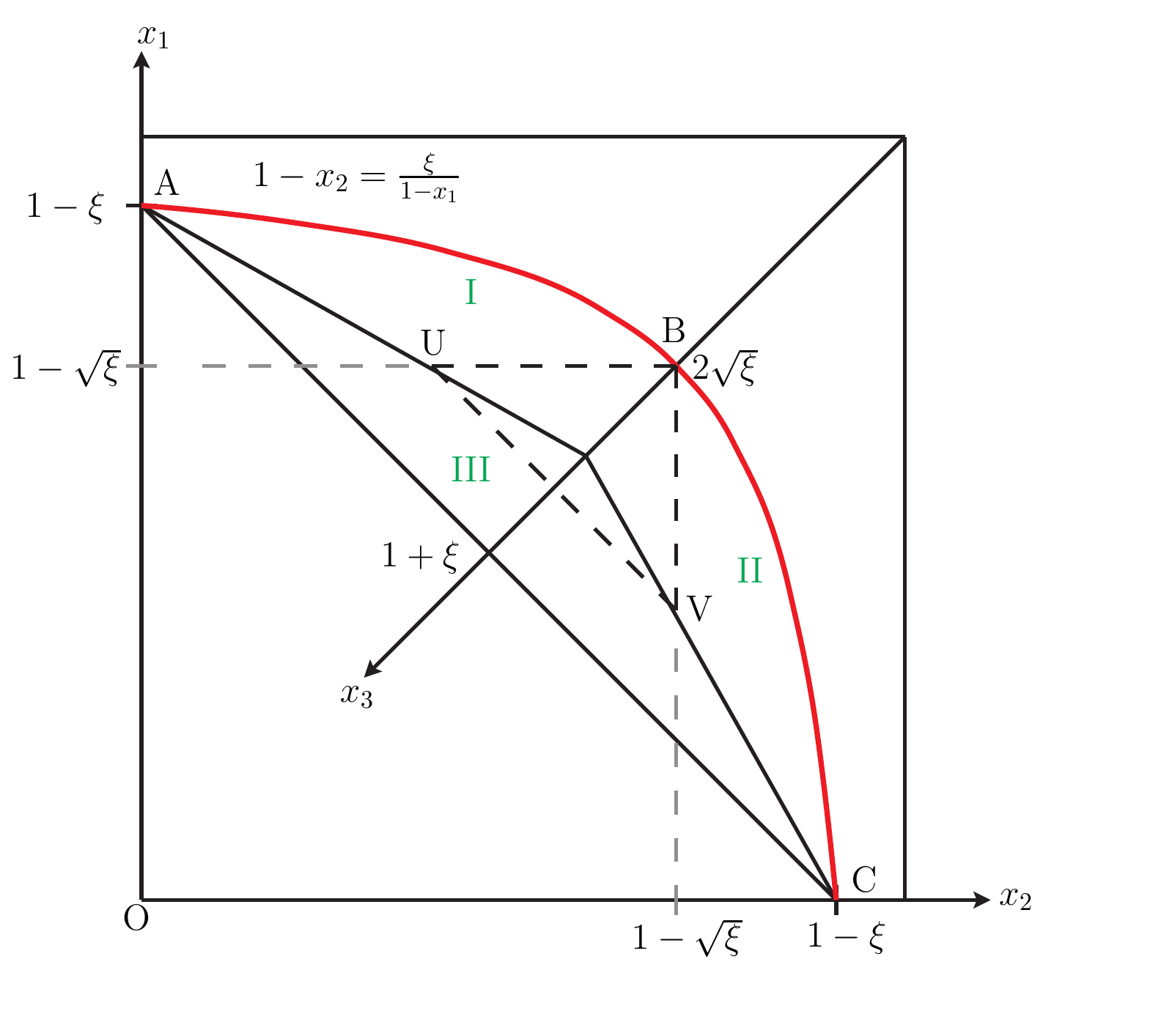} \caption{Dalitz plot showing phase space for $\gamma\rightarrow q\bar{q}g$
with off-shell gluon. The energy momentum conservation condition $x_{1}+x_{2}+x_{3}=2$
is satisfied throughout this $x_{1}-x_{2}$ plane and the actual length
along $x_{3}$ axis is $\sqrt{2}$ times the measured length. The
collinear limit (when the gluon is collinear to the quark) corresponds
to $x_{1}=1-\xi$, $x_{2}=0$, while the soft limit (when the gluon
is soft to the quark) corresponds to $x_{1}=x_{2}=1-\sqrt{\xi}$.
The soft boundary of the phase space $1-x_{2}=\xi/(1-x_{1})$ is denoted
by the red curve. }
\label{dalitz2} 
\end{figure}

As discussed above we need to construct the characteristic function
$\ensuremath{\mathcal{F}}(e,\xi)$ for these three event shape variables
for the order $\alpha_{s}$ process $\gamma^{*}\rightarrow q\bar{q}g$.
The color stripped squared matrix element after removing the coupling
is 
\begin{align}
\mathcal{M}(x_{1},x_{2},\xi)=\frac{(x_{1}+\xi)^{2}+(x_{2}+\xi)^{2}}{(1-x_{1})(1-x_{2})}-\frac{\xi}{(1-x_{1})^{2}}-\frac{\xi}{(1-x_{2})^{2}},
\end{align}
where the energy fractions are defined by 
\begin{align}
x_{1}= & \frac{2p_{1}\cdot Q}{Q^{2}},\quad x_{2}=\frac{2p_{2}\cdot Q}{Q^{2}},\quad x_{3}=\frac{2k\cdot Q}{Q^{2}}.\label{energyfrac}
\end{align}
Here $k$ denotes the momentum of the off-shell gluon and, $p_{1}$
and $p_{2}$ are the momenta of the quark and anti-quark respectively.
Momentum conservation $Q=p_{1}+p_{2}+k$ gives the constraint $x_{1}+x_{2}+x_{3}=2$.
Figure (\ref{dalitz2}) gives the Dalitz plot for this processes. 
In the soft gluon limit we approximate the squared matrix element
to, 
\begin{align}
\mathcal{M}_{{\rm {soft}}}(x_{1},x_{2},\xi)=\frac{2}{(1-x_{1})(1-x_{2})}.\label{softmatr}
\end{align}
Note that this is the same as what we would write in the soft gluon
limit for the case of massless gluon.

Next we will take the mentioned three shape variables in turn and
construct eikonalized versions of them and then compute 
the corresponding characteristic functions, and their Borel functions.
\subsection{Thrust}
%
%%%%%%%%%%%%%%%%%%%%%%%%%%%%%%
Thrust is one of the most studied event shapes and it has a historical
connection with the determination of strong coupling constant $\alpha_{s}$.
It is defined as~\cite{Farhi:1977sg} 
\begin{align}
T=\underset{{\bf {n}}}{\text{Max}}\,\,\frac{\sum_{i}|{\bf p}_{i}\cdot{\bf {n}}|}{\sum_{i}E_{i}},\label{thrustdef}
\end{align}
where ${\bf p}_{i}$ denotes the 3-momentum of the $i$-th particle
in the final state and \textbf{{n}} is a unit vector. %%%%%%%%%%%%%%%%%%%%%%%%%%%%%%%
In order to determine the range of $T$, we need to consider two extreme
cases: a most spherical configuration and a pencil like configuration.
For a spherical configuration, $T$ attains a minimum value $1/2$
and for a pencil like configuration, $T$ attains a maximum value
$1$. Thus, thrust varies in the range $1/2\leq T\leq1$. For a three
particles final state, the numerator in Eq. (\ref{thrustdef}) is maximum
when ${\bf {n}}$ is along the direction of the largest ${\bf {p_{i}}}$.
Thus, the thrust for all massless particles final state is given by
\begin{align}
T=\text{Max}\lbrace x_{1},x_{2},x_{3}\rbrace.\label{thrustredef}
\end{align}
In presence of a massive off-shell gluon in the final state, the definition
of thrust needs some simple modifications which was first given in~\cite{Gardi:1999dq}
and has the form, 
\begin{align}
T=\text{Max}\left\lbrace x_{1},x_{2},\sqrt{x_{3}^{2}-4\xi}\right\rbrace .\label{thrustmass}
\end{align}
Substituting in the definition Eq. (\ref{charf}) of the characteristic
function, the squared matrix element Eq. (\ref{softmatr}) and the definition
of thrust Eq. (\ref{thrustmass}), we obtain 
\begin{align}
\mathcal{F}(T,\xi)=\int\int dx_{1}\ dx_{2}\frac{2}{(1-x_{1})(1-x_{2})}\delta\left(T-\text{Max}\left\lbrace x_{1},x_{2},\sqrt{x_{3}^{2}-4\xi}\right\rbrace \right).\label{charf2}
\end{align}
When the radiated (dressed) gluon is soft, this integral receives
contributions from the regions $\text{I}$ and $\text{II}$ as shown
in Figure (\ref{dalitz2}). Region $\text{I}$ contributes when $x_{1}$
is the largest, and region $\text{II}$ contributes when $x_{2}$
is the largest of $x_{1},x_{2},\sqrt{x_{3}^{2}-4\xi}$. Naming these
contributions as $\mathcal{F}_{1}(T,\xi)$ and $\mathcal{F}_{2}(T,\xi)$
respectively, we have 
\begin{align}
\mathcal{F}(T,\xi)\simeq\mathcal{F}_{1}(T,\xi)+\mathcal{F}_{2}(T,\xi),\label{feqn}
\end{align}
where 
\begin{align}
\mathcal{F}_{1}(T,\xi)= & \int_{2-T-\sqrt{T^{2}+4\xi}}^{\frac{1-\xi-T}{1-T}}dx_{2}\ \mathcal{M}(T,x_{2},\xi)\nonumber \\
\nonumber \\
\mathcal{F}_{2}(T,\xi)= & \int_{2-T-\sqrt{T^{2}+4\xi}}^{\frac{1-\xi-T}{1-T}}dx_{1}\ \mathcal{M}(x_{1},T,\xi),\label{int1}
\end{align}
Note that the integrals $\mathcal{F}_1$ and $\mathcal{F}_2$ are same due to the symmetry of $\mathcal{M}(x_{1},x_{2},\xi)$ under
interchange of $x_1$ and $x_2$.
The limits of the integration are determined by the boundary of the
phase space shown in red in the Figure (\ref{dalitz2}). The characteristic
function immediately evaluates to 
\begin{align}
\mathcal{F}\left(t,\xi\right)=-\frac{4}{t}\text{log}\bigg(\frac{\xi}{t(q-t)}\bigg),\label{Ft2}
\end{align}
where $t\equiv1-T$ and $q=\sqrt{T^{2}+4\xi}$. Now, using Eqs. (\ref{borfu})
and (\ref{Ft2}) the Borel function for the thrust takes the form, 
\begin{align}
B(t,u)= & \frac{4\sin\pi u}{\pi u}\frac{1}{t}e^{\frac{5u}{3}}\int_{t^{2}}^{t}\xi^{-u-1}d\xi,
\end{align}
where the lower limit is determined using the collinear gluon boundary
conditions, $x_{1}=1-\xi$, $x_{2}=0$ and the upper limit is determined
from the soft gluon boundary condition $x_{1}=x_{2}=1-\sqrt{\xi}$.
Evaluating the integral we immediately obtain 
\begin{align}
B(t,u)= & \frac{\sin\pi u}{\pi u}e^{\frac{5u}{3}}\frac{4}{u}\frac{1}{t}\bigg(\frac{1}{t^{2u}}-\frac{1}{t^{u}}\bigg),\label{borthrust}
\end{align}
this agrees with the leading singular terms of the same function presented
in~\cite{Gardi:2001ny}. Thus, it is possible to calculate the leading
singular terms in $\mathcal{F}\left(t,\xi\right)$ and $B(t,u)$ using
the eikonal matrix element.

\subsection{$C$-parameter}

\noindent \label{cpar2} The $C$-parameter was originally defined
in~\cite{Parisi:1978eg,Donoghue:1979vi} using the eigenvalues of
the matrix 
\begin{align}
\theta_{\alpha\beta}=\frac{1}{\sum_{j}|\textbf{p}^{(j)}|}\sum_{i}\frac{\textbf{p}_{\alpha}^{(i)}\textbf{p}_{\beta}^{(i)}}{|\textbf{p}^{(i)}|},
\end{align}
where $\textbf{p}_{\alpha}^{(i)}$ are the spatial component of the
momentum of $i$-th particle. If the eigenvalues of the above matrix
are denoted by $\lambda_{1}$, $\lambda_{2}$ and $\lambda_{3}$,
then the $C$-parameter is given by 
\begin{align}
C=3(\lambda_{1}\lambda_{2}+\lambda_{2}\lambda_{3}+\lambda_{1}\lambda_{3}).
\end{align}
This can be cast into a Lorentz invariant form 
\begin{align}
C=3-\frac{3}{2}\sum_{i,j}\frac{\left(p^{(i)}\cdot p^{(j)}\right)^{2}}{(p^{(i)}\cdot q)(p^{(j)}\cdot q)},\label{Cpardef}
\end{align}
where $p^{(i)}$ denotes the four momentum of the $i$-th particle
and $q$ denotes the total four-momentum. $C$ takes a minimum value
$0$ for a two-jet event and $C$ attains a maximum value $1$ for
a spherical event. 
If, however, the final state has a planar configuration the largest
value that the parameter can attain is $3/4$. This upper limit also
applies for the case of 3-body final state that concerns us. The above
expression of the $C$-parameter and its rescaled version can be written
down using the energy fractions and the virtuality of the off-shell
gluon. 
\begin{align}
c=\frac{C}{6}=\frac{(1-x_{1})(1-x_{2})(x_{1}+x_{2}-1+2\xi)-\xi^{2}}{x_{1}x_{2}(2-x_{1}-x_{2})}.
\end{align}
Now, we define the eikonalized version of the $c$-parameter 
\begin{align}
c_{eik}(x_{1},x_{2})=\frac{(1-x_{1})(1-x_{2})}{(1-x_{1})+(1-x_{2})},\label{cdef}
\end{align}
which coincides with the above definition in the soft gluon limit.
Note that $c_{eik}$ is not a function of the virtuality $\xi$. We
will use $c_{eik}$ to calculate the characteristic function for $C$-parameter;
it is convenient to change variables from $x_{1}$ and $x_{2}$ into
$y=2-x_{1}-x_{2}$ and $z=(1-x_{2})/y$. In these new variables $c_{e}ik(y,z)=yz(1-z)$.
The characteristic function (Eq. (\ref{charf})) in this limit takes
the form, 
\begin{align}
\mathcal{F}=\int dy\ dz\ y\ \mathcal{M}_{\text{soft}}(y,z,\xi)\ \delta(c_{eik}(y,z)-c),\label{Fcpar}
\end{align}
where, 
\begin{align}
\mathcal{M}_{\text{soft}}(y,z,\xi)=\frac{2}{y^{2}z(1-z)}
\end{align}
The symmetry of $\mathcal{M}$ under $x_{1}\leftrightarrow x_{2}$
appears as symmetry under $z\rightarrow1-z$. In order to perform
the integral in Eq. (\ref{Fcpar}), it is required to determine the
limits of the $z$-integration using the boundary of the soft region
that is given by $x_{2}=(1-\xi-x_{1})/(1-x_{1})$. The integral in
Eq. (\ref{Fcpar}) has an explicit form, 
\begin{align}
\begin{split}\mathcal{F}=\int_{2\sqrt{\xi}}^{1+\xi}dy & \int_{\frac{1}{2}-\frac{1}{2}\sqrt{1-4\xi/y^{2}}}^{\frac{1}{2}+\frac{1}{2}\sqrt{1-4\xi/y^{2}}}dz\,\,\frac{2}{yz(1-z)}\frac{1}{y\sqrt{1-4c/y}}\,\,\Big(\delta(z-z_{1})+\delta(z-z_{2})\Big),\end{split}
\label{Fcpar2}
\end{align}
where 
\begin{align}
z_{1}=\frac{1}{2}+\frac{1}{2}\sqrt{1-4c/y}\quad\text{and}\quad z_{2}=\frac{1}{2}-\frac{1}{2}\sqrt{1-4c/y}.
\end{align}
This integral has a symmetry under $z\leftrightarrow(1-z)$ interchange,
therefore the integral over $z$ equals twice the integral between
$z=1/2$ and the upper limit in Eq. (\ref{Fcpar2}), where only the
$\delta(z-z_{1})$ is relevant. The condition $z\leq\frac{1}{2}+\frac{1}{2}\sqrt{1-4\xi/y^{2}}$
implies that $y\geq\xi/c$. With this the integral in Eq. (\ref{Fcpar2})
takes the form 
\begin{align}
\mathcal{F}=\frac{4}{c}\int_{\xi/c}^{1+\xi}dy\,\,\frac{1}{y\sqrt{1-4c/y}}.
\end{align}
Evidently, it is only the lower limit of the integration that gives
rise to singular contribution in the $\xi\rightarrow0$ limit. 
As we are only interested in the derivative of ${\cal F}$, we get,
without even evaluating the integral 
\begin{align}
\frac{d\mathcal{F}}{d\xi}=-\frac{4}{c\xi}\frac{\sqrt{\xi}}{\sqrt{\xi-4c^{2}}}.
\end{align}
Contrast this to the computation of $d{\cal F}/{d\xi}$ presented
in \cite{Gardi:2003iv} where the computation proceeds with the full
definition of the $c$-parameter. In that paper the authors had to
deal with the complicated elliptic integrals and had to carefully
consider small $c$ and small $\xi$ limits. These complications are
completely absent in our method.

Now we are in the position to compute the Borel function $B(c,u)$
for the $c$-parameter. We have to substitute $d{\cal F}/{d\xi}$
into Eq. (\ref{borrep}), 
\begin{align}
B(c,u)=\frac{4\sin{\pi u}}{\pi u}\frac{1}{c}e^{\frac{5u}{3}}\int_{4c^{2}}^{\frac{c}{1-c}}d\xi~\frac{\xi^{-u}}{\sqrt{\xi(\xi-4c^{2})}}
\end{align}
where the lower limit in the above integral is determined using $x_{1}=x_{2}=1-\sqrt{\xi}$
(soft limit), and the upper limit is determined using $x_{1}=1-\xi$,
$x_{2}=0$ (collinear limit). We are interested in the logarithmically
enhanced terms, thus we can replace the upper limit of the integration
by $c/(1-c)\approx c$. Carrying out the integral yields the Borel
function 
\begin{align}
B(c,u)=4\frac{\text{sin}\pi u}{\pi u}e^{\frac{5u}{3}}\frac{1}{c}\bigg[\frac{1}{(2c)^{2u}}\frac{\sqrt{\pi}\Gamma(u)}{\Gamma(u+\frac{1}{2})}-\frac{1}{uc^{u}}\bigg].
\end{align}
Our result agrees with the soft contribution of the same function
presented in~\cite{Gardi:2003iv}. 
We conclude thus, that the leading singular terms in $\mathcal{F}(c,\xi)$
and $B(c,\xi)$ can be captured with significant ease if we use the
eikonal version $c_{eik}$ that we have introduced for the $c$-parameter.
\subsection{Angularities}

\noindent \label{ang2} As a demonstration of the wide applicability
of our method we consider one more event shape variable -- the angularities.
Angularities are novel observables that allow us to transform between
recoil-insensitive to recoil-sensitive observables in a continuous
manner. Angularities were first introduced almost twenty years ago
in~\cite{Berger:2003pk,Berger:2003iw,Berger:2002ig} and they were
defined as 
\begin{align}
\tau_{a}=\frac{1}{Q}\sum_{i}E_{i}(\sin\theta_{i})^{a}(1-|\cos\theta_{i}|)^{1-a},\label{angdef}
\end{align}
where $\theta_{i}$ is the angle made by $i$-th particle with the
thrust axis, $E_{i}$ is the energy of the particle $i$ and $a$
is a continuous parameter. The thrust axis is defined as the axis
with respect to which Eq. (\ref{angdef}) is minimized at $a=0$. One
can easily realize that angularities with $a=0$ corresponds to $1-T$,
where $T$ is the thrust, while $a=1$ refers to jet broadening \cite{Catani:1992jc}.
The continuous parameter $a$ has a range $-\infty<a<2$, where the
upper limit on $a$ is fixed by infrared safety. 
In terms of $x_{i}$ and $\xi$ angularities were defined in \cite{Berger:2004xf}
as, 
\begin{align}
\tau_{a}(x_{1},x_{2},\xi)= & \frac{1}{x_{1}}(1-x_{1})^{1-a/2}\bigg[(1-x_{2}-\xi)^{1-a/2}(x_{1}+x_{2}-1+\xi)^{a/2}\nonumber \\
 & +(1-x_{2}-\xi)^{a/2}(x_{1}+x_{2}-1+\xi)^{1-a/2}\bigg],\label{angredef}
\end{align}
where, thrust axis is considered along $\textbf{p}_{1}$ (quark momentum).
As we did for the $c$-parameter we introduce an eikonal version of
the angularities: 
\begin{align}
\tau_{a}^{eik}(x_{1},x_{2},\xi)=(1-x_{1})^{1-a/2}(1-x_{2})^{a/2}.\label{Ftaupar}
\end{align}
Now, Using Eq. (\ref{charf}) and (\ref{softmatr}) the characteristics
function takes the form, 
\begin{align}
\mathcal{F}=\int dx_{1}dx_{2}\frac{2}{(1-x_{1})(1-x_{2})}\delta\big(\tau_{a}^{eik}(x_{1},x_{2},\xi)-\tau_{a}\big).\label{Fa2}
\end{align}
It is straight-forward to perform the $x_{1}$ integration to obtain
% and the characteristics function takes the form, 
\begin{align}
\mathcal{F} & =\frac{4}{\tau_{a}(1-\frac{a}{2})}\int dx_{2}\,\frac{1}{1-x_{2}}.\label{Ftauw}
\end{align}
We determine the upper limit of this integration using the soft boundary
$1-x_{2}=\xi/(1-x_{1})$. As shown in \cite{Berger:2004xf}, the lower
limit of this integration does not contribute to the logarithmically
enhanced terms. The upper limit of the integration is 
\begin{align*}
1-\bigg(\frac{\xi^{1-\frac{a}{2}}}{\tau_{a}}\bigg)^{\frac{1}{1-a}}.
\end{align*}
We finally have the characteristics function 
\begin{align}
\mathcal{F}(\tau_{a},\xi)=-\frac{4}{\tau_{a}}\frac{1}{1-a}\log\xi.\label{Ftaupar3}
\end{align}
Taking the derivative with respect to $\xi$ and substituting in Eq.
(\ref{borfu}) we get the Borel function %Using Eq. \ref{Ftaupar3} we obtain 
\begin{align}
B(\tau_{a},u)=\frac{4\sin{\pi u}}{\pi u}\frac{1}{1-a}\frac{1}{\tau_{a}}e^{\frac{5u}{3}}\int_{\tau_{a}^{2}}^{\tau_{a}^{\frac{2}{2-a}}}\hspace{-2mm}d\xi~\xi^{-u-1},\label{uyt}
\end{align}
where the limits are determined using the collinear and soft gluon
boundary conditions mentioned in Figure (\ref{dalitz2}). Upon performing
the integration in Eq. (\ref{uyt}) we obtain 
\begin{align}
B(\tau_{a},u)=\frac{\sin{\pi u}}{\pi u}e^{\frac{5}{3}u}\frac{4}{1-a}\frac{1}{u}\frac{1}{\tau_{a}}\left[\frac{1}{\tau_{a}^{2u}}-\frac{1}{\tau_{a}^{\frac{2u}{2-a}}}\right]\,,\label{Bau}
\end{align}
which agrees with the soft contribution of the same function presented
in~\cite{Berger:2004xf}. We have, thus obtained, the leading singular
terms in $\mathcal{F}(\tau_{a},\xi)$ and $B(\tau_{a},\xi)$, which
are responsible for power corrections by considering the eikonal matrix
element and the eikonal version of the angularities $\tau_{a}^{eik}(x_{1},x_{2},\xi)$
which again substantially simplifies the computation. 

\section{Borel function using Eikonal Dressed Gluon Exponentiation \\ in the light-cone variables}

\label{EDGE} In this section, we will follow the same steps of Sec. (\ref{EDGEenergy}) and calculate Borel function for thrust, $C$-parameter
and angularities using a different set of kinematic variables. Instead
of the energy fractions that we used in the previous section we would
employ the transverse momentum $k_{\perp}$ and rapidity $y$ of the
massive eikonal gluon. 
In the soft gluon approximation, a number of event shapes for massless
particles were defined in ~\cite{Salam:2001bd,Mateu:2012nk}. We
will consider a class of event shapes which, for massive soft gluon
emission, can be written as %\be
\begin{align}
\bar{e}(k,Q)\,=\,\sqrt{\frac{k_{\perp}^{2}+k^{2}}{Q^{2}}}~h_{e}(y)\,,\label{class}
\end{align}
where $k_{\perp}$ and $y$ denote transverse momentum of the gluon
and pseudo-rapidity measured with respect to the thrust axis respectively.
The function $h_{e}(y)$ characterizes the given event shape. Some
of the approximations described below apply to more general event
shapes as well, but the results are especially simple for those which
can be cast in the form of Eq. (\ref{class}).

The contribution from the emission of a soft off-shell gluon can easily
be computed applying the eikonal approximation to the vertex for the
emission from the hard parton. Since off-shell soft-gluon phase space
factorizes~\cite{Gardi:2001di} from the hard partons, and also the
matrix element factorizes, 
the soft cross section takes on a simple and universal form, 
\begin{equation}
\frac{d\sigma}{\sigma}\,=\,\frac{1}{3}\,\frac{4}{k^{2}+k_{\perp}^{2}}\,dk_{\perp}^{2}dy\,.
\end{equation}
The characteristic function is also then given, in the soft limit,
by a simple and universal expression 
\begin{equation}
\ensuremath{\mathcal{F}}(e,\xi)=\int dk_{\perp}^{2}dy\,\frac{2}{k^{2}+k_{\perp}^{2}}\delta\left(e-\bar{e}\left(k^{2},k_{\perp}^{2},y\right)\right)\,,\label{uniF}
\end{equation}
which integrates to the remarkably simple form, 
\begin{equation}
\ensuremath{\mathcal{F}}(e,\xi)\,=\,\frac{8}{e}\int_{y_{{\rm min}}}dy\,\label{uniF2}
\end{equation}
where the only information on the chosen observable is the phase space
boundary given by the minimum rapidity $y_{{\rm min}}$. The upper
limit of integration is not relevant, since it does not give any singular
contributions in the $\xi\rightarrow0$ limit, which is the only significant
limit for power corrections.

Up to now, we have kept the discussion generic, for any shape belonging
to the class given in Eq. (\ref{class}). Let us now illustrate the
results by looking at some specific examples. \\

\subsection{Thrust}

\label{thrustedge} The thrust for a generic process is defined in
Eq. (\ref{thrustredef}). In the two jet events all event shape variables
that we consider tends to $0$, except thrust which tends to 1. Thus,
it is convenient to define $t=1-T$. In the soft gluon approximation,
thrust in terms of the $k_{\perp}$ and rapidity $y$ is given by
\cite{Salam:2001bd} 
\begin{align}
t=\frac{1}{Q}\sqrt{k_{\perp}^{2}+k^{2}}\ e^{-|y|}.\label{thrustrapid}
\end{align}
Note that, for our case the gluon is massive and $k^{2}\neq0$. 
In order to perform the integral in Eq. (\ref{uniF2}), we need to determine
the lower limit of the rapidity. The lower limit of rapidity $y$
is determined by putting $k_{\perp}=0$ in Eq. (\ref{thrustrapid}),
thus minimum rapidity is given by, 
\begin{equation}
y_{{\rm min}}\,=\,\,\ln\Big(\frac{1}{t}\sqrt{\xi}\Big).\label{minythrust}
\end{equation}
Now, using Eq. (\ref{uniF2}) and (\ref{minythrust}) the characteristics
function has the form, 
\begin{align}
\mathcal{F}=-\frac{8}{t}\text{log}(\frac{\sqrt{\xi}}{t}).
\end{align}
The Borel function $B(t,u)$ is then given by 
\begin{equation}
B(t,u)
=\frac{\sin{\pi u}}{\pi u}e^{\frac{5}{3}u}\frac{4}{u}\frac{1}{t}\left[\frac{1}{t^{2u}}-\frac{1}{t^{u}}\right]\,,\label{Bau}
\end{equation}
which is in well agreement with the soft approximated version of the characteristics
function and Borel function presented in \cite{Gardi:2001ny}.

\subsection{$C$-parameter}

\noindent The $C$-parameter for a generic process is defined in Eq.
(\ref{cpar2}). The $C$-parameter in the soft approximation and expressed
using $k_{\perp}$ and $y$ is given by \cite{Salam:2001bd} 
\begin{align}
c=\frac{C}{6}=\frac{1}{2Q}\sqrt{k^{2}+k_{\perp}^{2}}\ \frac{1}{\text{cosh}\ y}.\label{crapiddef}
\end{align}
As for the case of thrust %As mentioned in sec. \ref{thrustedge}, 
we determine the lower limit of rapidity by putting $k_{\perp}=0$
and obtain 
\begin{align}
y_{{\rm min}}=\cosh^{-1}\left({\sqrt{\xi}}/(2c).\right).
\end{align}
Now, substituting $y_{\text{min}}$ in Eq. (\ref{uniF2}) we obtain
the characteristic function in the soft gluon limit: 
\begin{align}
\mathcal{F}=-\frac{8}{c}\text{cosh}^{-1}\bigg(\frac{\sqrt{\xi}}{2c}\bigg).\label{Fcpar4}
\end{align}
This yields the Borel function %\be
\begin{equation}
B(c,u)
=4\frac{\text{sin}\pi u}{\pi u}e^{5u/3}\frac{1}{c}\bigg[\frac{1}{(2c)^{2u}}\frac{\sqrt{\pi}\Gamma(u)}{\Gamma(u+\frac{1}{2})}-\frac{1}{uc^{u}}\bigg],\label{Bcu}
\end{equation}
in full agreement with the soft contribution to the same function
in~\cite{Gardi:2003iv}. Notice that, while collinear effects present
in~\cite{Gardi:2003iv} are not properly reproduced, as expected,
the cancellation of the pole at $u=0$, which is a consequence of
the IR safety of the event shape, is preserved.

\subsection{Angularities}
In the soft gluon limit, angularities takes the form \cite{Salam:2001bd,Mateu:2012nk},
\begin{align}
\tau_{a}=\frac{1}{Q}\sqrt{k^{2}+k_{\perp}^{2}}\ e^{-|y|(1-a)},
\end{align}
and the minimum rapidity is given by 
\begin{equation}
y_{{\rm min}}\,=\,\frac{1}{1-a}\,\ln\Big(\frac{1}{\tau_{a}}\sqrt{\xi}\Big)\,.\label{minyang}
\end{equation}
Now, using Eq. (\ref{uniF2}), one easily finds 
\begin{equation}
\frac{d\ensuremath{\mathcal{F}}(\tau_{a},\xi)}{d\xi}\,=\,-\frac{1}{1-a}\,\frac{4}{\tau_{a}\xi}\,.\label{charfa}
\end{equation}
The Borel function $B(\tau_{a},u)$ is then given by 
\begin{equation}
B(\tau_{a},u)
=\frac{\sin{\pi u}}{\pi u}e^{\frac{5}{3}u}\frac{4}{1-a}\frac{1}{u}\frac{1}{\tau_{a}}\left[\frac{1}{\tau_{a}^{2u}}-\frac{1}{\tau_{a}^{\frac{2u}{2-a}}}\right]\,,\label{Bau}
\end{equation}
again in agreement with the soft contribution to the results of~\cite{Berger:2004xf},
and reproducing, in the limit $a\to0$, the results for thrust of
Ref.~\cite{Gardi:2002bg}.
\section{The Sudakov exponent \label{sec:Exponentiation}}
In this section, we will describe the computation of Sudakov exponent
for thrust. Similar conclusions hold true for the other two shape variables as well that we 
have considered in this article. We can calculate the Borel function in the Laplace space $B_{\nu}^{t}(u)$
in the eikonal limit using $B(t,u)$ that we wrote above in Eq.~(\ref{borthrust}), we obtain,
\begin{align}
B_{\nu}^{t,\text{eik}\text{ }}(u) & =\frac{4\,\text{sin}\,\pi u}{\pi u}\,e^{\frac{5u}{3}}\,\frac{1}{u}\,\bigg[\bigg(\nu^{2u}\gamma(-2u,\nu)+\frac{1}{2u}\bigg)-\bigg(\nu^{u}\gamma(-u,\nu)+\frac{1}{u}\bigg)\bigg],\label{eq: borel-thrust}
\end{align}
where, we have used
\begin{align}
\int_{0}^{1}\frac{dt}{t}\,e^{u\,\text{log}\frac{1}{t}}\,(e^{-\nu t}-1)=\nu^{u}\gamma(-u,\nu)+\frac{1}{u},\label{eq: integration-formula-1}
\end{align}
and $\gamma(-u,\nu)=\Gamma(-u)-\Gamma(-u,\nu)$. In the Sudakov region
($\nu\rightarrow\infty$), we can replace $\gamma(-u,\nu)$ by $\Gamma(-u,\nu)$.
Retaining only the logarithmically enhanced terms (powers
of $\text{\text{log\, }\ensuremath{\nu}}$), 
in the small $u$ limit  $B_{\nu}^{t}(u)$ takes the form,
\begin{align}
B_{\nu}^{t,\text{{eik}}}(u)=2\,e^{\frac{5}{3}u}\frac{\,\text{sin\,}\pi u}{\pi u}\bigg[\Gamma(-2u)\:\bigg(\nu^{2u}-1\bigg)\frac{2}{u}-\Gamma(-u)\:\bigg(\nu^{u}-1\bigg)\frac{2}{u}\bigg] & .\label{eq: borel-thrust-fin}
\end{align}
The first term inside the square brackets corresponds to large-angle
soft gluon emissions and the second term to collinear gluon emissions.
Note that this expression is free from any $u=0$ singularities.
There are two sources of the poles in this expression: $\Gamma(-2u)$ has poles for all positive
integers and half-integers, and $\Gamma(-u)$ has poles for all positive
integers. However, the pre-factor $\text{sin\,\ensuremath{\pi u}}$
regulates the poles at the integer values of $u$. Thus, $B_{t}^{\nu,\text{eik}}$
has renormalon poles only at half-integer values of $u$.

We will now compare our result with the full result for \textbf{$B_{\nu}^{t}$}
presented in \cite{Gardi:2001ny,Gardi:2003iv} which is
given by, 
\begin{align}
B_{\nu}^{t}(u)=\,2{\rm e}^{\frac{5}{3}u}\,\frac{\sin\pi u}{\pi u}\left[\Gamma(-2u)\left(\nu^{2u}-1\right)\frac{2}{u}-\Gamma(-u)\left(\nu^{u}-1\right)\left(\frac{2}{u}+\frac{1}{1-u}+\frac{1}{2-u}\right)\right]\,.\label{eq: Btfull}
\end{align}
Note the poles at $u=1$ and
$u=2$ which are absent in the collinear term of our eikonalized  $B_{\nu}^{t,\text{eik}}(u)$.
We further notice that no spurious renormalon poles are present in the eikonalized version. Recall that, for thrust
approximation was done only for the matrix element and not for the definition of the variable. 
To show that our eikonal versions of the shape variables do not spoil the above feature  we present the results for the $c$-parameter. The eikonal version and full result~\cite{Gardi:2003iv} are as follows:
\begin{align}
B_{\nu}^{c,\text{eik}}(u) & =2 \, \frac{\sin\,\pi u}{\pi u}e^{\frac{5u}{3}}\, \bigg[\Gamma(-2u)(\nu^{2u}-1)2^{1-2u}\frac{\sqrt{\pi}\Gamma(u)}{\Gamma(u+\frac{1}{2})}-\frac{2}{u}\Gamma(-u)(\nu^{u}-1)\bigg]\label{eq:ceikonal},   \\
%\end{align} 
%\begin{align}
B_\nu^c (u) & = 2  \,
\frac{\sin \, \pi u}{\pi u} e^{\frac{5u}{3}}\, \left[\Gamma(- 2 u) \left(\nu^{2 u} - 1 
\right) 2^{1 - 2 u} \frac{\sqrt{\pi} \Gamma(u)}{\Gamma(\frac12 + u)}
\right.\nonumber \\ 
& \hspace*{115pt} - \left. \Gamma(- u) \left({\nu}^{u} - 1 \right)
\left(\frac2u + \frac1{1 - u} + \frac{1}{2 - u} \right) \right]\,.
\label{Bnu}
\end{align}
Again, no spurious poles are introduced.
As expected, the EDGE does not reproduce 
the collinear renormalon singularities as it cannot capture the hard-collinear emissions correctly.

The perturbative
coefficients of the Sudakov exponent in the large-$\beta_0$ limit can be determined by
expanding $B_{\nu}^{t}(u)$ in powers of $u$ and replacing $u^n$ by
$n!/(\beta_{0}\alpha_{s}/\pi)^{n+1}$. We notice that the large-angle
soft gluon emission terms -- the coefficient of $\Gamma(-2u)$, are identical
in the eikonalized and full versions of the Borel function in the Laplace space.
%$B_{\nu}^{t,\text{eik}}(u)$ and $B_{\nu}^{t}(u)$.
 This implies that the leading logs -- the terms of the form $L^{n+1}\alpha_{s}^{n}$ where $L=\text{log \ensuremath{\nu} }$,
will be the same between the two.
The differences in the sub-leading logarithms appear due to the absence
of the $u=1$ and $u=2$ poles in the collinear terms.
We will now expand the two functions
to the first few
orders to demonstrate the matching of the LL terms and the discrepancy
in the sub-leading logarithms. The expansion of the full result gives,
\begin{alignat}{1}
B_{\nu}^{t}\left(u\right) & =-2\,L^{2}+0.691\,L\nonumber \\
 & +\,\left(-2\,L^{3}-5.297\,L^{2}-6.485\,L\right)\,u\hspace{-6pt}\nonumber \\
 & +\,\left(-1.167\,L^{4}-5.527\,L^{3}-14.491\,L^{2}-31.655\,L\right)\,u^{2}\nonumber \\
 & +\,\left(-0.5\,L^{5}-3.262\,L^{4}-12.329\,L^{3}-39.003\,L^{2}-80.940\,L\right)\,u^{3}\nonumber \\
 & +\,\left(-0.172\,L^{6}-1.405\,L^{5}-6.832\,L^{4}-28.452\,L^{3}-87.21\,L^{2}-175.80\,L\right)\,u^{4}\label{eq: Btexpan}\\
 & +\mathcal{O}(u^{5})+\ldots\nonumber 
\end{alignat}
whereas, the expansion of the eikonal result gives,
\begin{alignat}{1}
B_{\nu}^{t,\text{eik}}(u)= & -2\:L^{2}-2.31\:L\nonumber \\
+\: & (-2\,L^{3}-6.79\,L^{2}-15.71\,L)\:u\nonumber \\
+\, & (-1.167\,L^{4}-6.02\,L^{3}-19.10\,L^{2}-44.59\,L)\:u^{2}\nonumber \\
+\, & (-0.5\,L^{5}-3.38\,L^{4}-13.86\,L^{3}-45.47\,L^{2}-93.66\,L)\:u^{3}\nonumber \\
+\, & (-0.172\,L^{6}-1.429\,L^{5}-7.216\,L^{4}-30.61\,L^{3}-93.57\,L^{2}-187.395\,L)\:u^{4}\label{eq: Beikexpan}\\
+\, & \mathcal{{O}}(u^{5})+\ldots\:\,.\nonumber 
\end{alignat}
As expected, the leading logarithms are appearing correctly in the
eikonal approximated version of the Borel function in the Laplace
space. We observe that the differences in NLL and NNLL terms between
$B_{\nu}^{t}$ and $B_{\nu}^{t,\text{eik}}$ are decreasing as we
go higher order in $u$.

Let us now  discuss the power corrections. The Sudakov exponent is an integral over
$u$ and the poles of $B_{\nu}^{t}$ that occur on the real $u$-axis make it an ill defined 
integral. The integral can be defined by shifting the poles above or below the axis or equivalently
indenting the contour below or above the poles. This however, introduces an ambiguity that is proportional
to the residue of the poles. 
The poles of  $B_{\nu}^{t}$ that occur at $u=m/2$, where $m$ is an odd integer
give the ambiguity \cite{Gardi:2003iv} originating from the  large-angle
soft gluon emissions. 
From Eq.~(\ref{eq:sudakov-gen-1}) we see that the ambiguity would be of the form $(\Lambda \nu /Q)^m$
which implies the existence of non-perturbative power corrections of the same form. 
In Table (\ref{tab:soft-residues}) we present the residues of poles at
$u=m/2$ arising from $B_{\nu}^{t,\text{eik}}(u)$ which contribute
to the soft power corrections.
The full result $B_{\nu}^{t}(u)$ also has poles
at $u=1$ and $u=2$ which give indications to the size of the collinear
power corrections whereas these are absent in  $B_{\nu}^{t,\text{eik}}(u)$. 
%We will show the residues for first few powers of $\nu$ to show explicitly how the collinear
%corrections are suppressed as compared to the soft corrections. 
Thus, the collinear power corrections
exist only for $\nu^{1}$ and $\nu^{2}$, and using the full expression
for $B_{\nu}^{t}(u)$ we find that they are given by $-2\big(\frac{\overline{\Lambda}}{Q}\big)^{2}$
and $-\frac{1}{2}\big(\frac{\overline{\Lambda}}{Q}\big)^{4}$ respectively.
Note that, in the calculation of the residues for the collinear terms
we have ignored the $\mathcal{O}(1)$ factor $C_{F}/2\beta_{0}$. We see, thus, that 
the residue of the collinear power correction is suppressed by $\bar{\Lambda}/Q$
as compared to the soft correction for $\nu^{1}$ term. For example at the LEP where $Q=209$ Gev and $\bar{\Lambda}=200\:e^{\frac{5}{6}}$
Mev, the ratio of the size of the collinear correction to the soft
correction for $\nu^{1}$ is approximately $-0.0017$. Thus, at colliders
like LEP, the soft power corrections are more important as compared
to the collinear corrections. As pointed out in \cite{Gardi:2001ny},
the dominant power correction arising from the residue at $u=1/2$ 
is proportional to $\nu$ and thus generates a shift in the resummed cross-section.

For the other event shapes considered in this article one can calculate
the Borel function in Laplace space using EDGE.
It remains true that  soft power corrections are dominant over the collinear corrections for all the event shapes considered in this paper.

\begin{table}[H]
\centering{}%
\begin{tabular}{|c|c|}
\hline 
Correction & Residue\tabularnewline
\hline 
$\nu^{1}$ & 8$\frac{\overline{\Lambda}}{Q}$\tabularnewline
\hline 
$\nu^{3}$ & $-\frac{4}{27}\big(\frac{\overline{\Lambda}}{Q}\big)^{3}$\tabularnewline
\hline 
$\nu^{5}$ & $\frac{1}{375}\big(\frac{\overline{\Lambda}}{Q}\big)^{5}$\tabularnewline
\hline 
$\nu^{7}$ & $-\frac{1}{30870}\big(\frac{\overline{\Lambda}}{Q}\big)^{7}$\tabularnewline
\hline 
\end{tabular}\caption{The size of the residues of renormalon singularities for soft power
corrections. The numbers quoted are $\pi$ times the residue and $\overline{\Lambda}=\Lambda\:e^{5/6}$.
We ignore here the $\mathcal{O}(1)$ factor $C_{F}/2\beta_{0}$ in
Eq. (\ref{eq:sudakov-gen-1}). \label{tab:soft-residues}}
\end{table}

\section{Conclusions}

In this paper, we have introduced Eikonal Dressed Gluon Exponentiation
which is a combination of Dressed Gluon Exponentiation and Eikonal
approximation. Using our method, we have demonstrated for several
event shapes at $e^{+}e^{-}$ colliders that the leading singular
contributions for the respective Borel functions in the single dressed
gluon approximation are produced with remarkably simple calculations.
It is straightforward to construct the Sudakov exponent in the large-$\beta_{0}$
limit for the power corrections using the procedure presented in ~\cite{Gardi:2001di,Gardi:1999dq}.
This exponentiation effectively resums both the large Sudakov logarithms
and the power corrections. We observe that EDGE does not introduce any spurious renormalons 
and correctly produces the dominant power corrections originating from soft emissions.

We have shown that in order to accurately capture the leading singular
terms of the characteristic function $\mathcal{F}(e,\xi)$ and Borel
function $B(e,\xi)$ for an event shape variable $e$, it is sufficient
to use the eikonal squared matrix element $\mathcal{M}$, together
with the eikonal version of the event shape variable. Typically the
shape variables such as $C$-parameter and angularities have complicated
expressions especially so when the final state gluon is off the mass
shell. We have demonstrated that the simplification of the computations
is achieved, both when one uses the energy fractions as the variables,
and also when we uses light-cone variables to parameterize the phase
space of the eikonal dressed gluon. When using the latter variables,
we observe that the minimum value of rapidity $y_{\text{min}}$ is
the source of the leading singular terms in $\mathcal{F}(e,\xi)$.
We believe that this method is sufficiently simple and flexible to
be implemented also in the more intricate environment of hadron collisions,
where hadronic event shapes and jet shapes provide important tools
for QCD analyses. 

\section{Acknowledgments}

SP and AM would like to thank MHRD Govt. of India for an SRF fellowship.
AT would like to thank Lorenzo Magnea for suggesting this project
and for very fruitful discussions, Einan Gardi for very useful discussions,
and the University of Turin and INFN Turin for warm hospitality during
the course of this work. \bibliographystyle{JHEP}
\bibliography{fullref}

\end{document}